\documentclass[cp1251
               ]{jetp} 

\begin{document}
\english
\newcommand{\WI}[2]{#1_{\mathrm{#2}}}
\newcommand{\isn}[2]{\mbox{$^{#2}${#1}}}

\title{Simulation of the LSD Response
to the Neutrino Burst from SN 1987A
} 




\author{K.~V.}{Manukovskiy} 
\email{manu@itep.ru}
\affiliation{NRC ``Kurchatov Institute'' --- Institute for Theoretical and Experimental Physics,
Moscow, 117218 Russia} 

\author{A.~V.}{Yudin} 
\email{yudin@itep.ru}
\affiliation{NRC ``Kurchatov Institute'' --- Institute for Theoretical and Experimental Physics,
Moscow, 117218 Russia}
\affiliation{NRC ``Kurchatov Institute'', Moscow, 123182 Russia}

\author{N.~Yu.}{Agafonova} 
\email{agafonova@inr.ru}
\affiliation{Institute for Nuclear Research, Russian Academy of Sciences, Moscow, 117312 Russia}

\authord{A.~S.}{Malgin} 
\affiliation{Institute for Nuclear Research, Russian Academy of Sciences, Moscow, 117312 Russia}

\authord{O.~G.}{Ryazhskaya} 
\affiliation{Institute for Nuclear Research, Russian Academy of Sciences, Moscow, 117312 Russia}

 \rtitle{Simulation of the LSD Response
 \dots 
 }

 \rauthor{Manukovskiy, Yudin, Agafonova et al.
 }

\abstract{Using the \textsc{Geant4} code, we have performed a full-scale simulation of the LSD response to the neutrino
burst from SN~1987A. The neutrino flux parameters were chosen according to one of the models: the
standard collapse model or the rotational supernova explosion model. We showed that, depending on the
chosen parameters, one can either obtain the required number of pulses in the detector or reproduce their
energy spectrum, but not both together. The interaction of neutrino radiation both with LSD itself and with
the material of the surrounding soil was taken into account in our simulation. We also explored the hypothesis
that the entire unique LSD signal at 2:52 UT was produced by neutron fluxes from the surrounding granite.
However, this hypothesis was not confirmed by our simulation. The results obtained provide a rich material
for possible interpretations.
}

\maketitle


\section{INTRODUCTION}
Supernova (SN) 1987A exploded on February 23,
1987, triggered a rapid development of the SN explosion
theory and experimental neutrino detection
methods. This SN was discovered in the optical range
\cite{IAUC4316} and then was observed in all ranges of the electromagnetic
spectrum. However, in addition, it was
detected by four neutrino detectors: the scintillation,
LSD \cite{Aglietta1987} and BUST \cite{Alekseev}, and Cherenkov, IMB \cite{Bionta} and
Kamiokande II (hereafter KII) \cite{Hirata}, ones. It should be
emphasized that this is the first and so far, unfortunately,
the sole case of neutrino detection from supernovae.

The LSD signal from SN 1987A recorded approximately
5~h before the IMB, KII, and BUST detector
signals still provokes debates about its origin. In the
universally accepted standard collapse model (a
spherically symmetric nonrotating star) \cite{MSK1,MSK2} the neutrino
radiation must be a single one, with a duration
$\sim 10$~s, and, therefore, the signal in LSD must be
observed simultaneously with the signals in the
remaining detectors. In addition, the number of pulses
in LSD must be smaller than that in IMB and KII,
because the masses of the IMB (5000~t of $\isn{H}{}_2\isn{O}{}$) and
KII (2140~t of $\isn{H}{}_2\isn{O}{}$) working material are greater than
the LSD mass (90~t of $\isn{C}{}_n\isn{H}{}_{2n}$). It is worth noting that
all these detectors were designed to detect electron
antineutrinos.

However, there are models \cite{Rujua1987, Berezinsky1988, Imshennik1995, Drago2016} that also admit a
double neutrino burst from the SN. As was shown in
\cite{ImshRya2004}, the signal at 2:52 UT recorded only in LSD and
containing five pulses can be explained in terms of the
rotating collapsar model \cite{Imshennik1995} as a result of the interaction
of high-energy electron neutrinos with the detector
structural elements that contained $\sim 170$~t of iron.

Besides, a similarity of the LSD signal energy characteristics
to the spectrum of gamma-ray photons
from neutron captures by iron nuclei, $n + \isn{Fe}{56} \rightarrow \isn{Fe}{57} + \gamma$ \cite{Yen}, engages our attention. Could the soil
surrounding the detector (Mont Blanc granite) irradiated
by neutrino fluxes from the SN become a source
of neutrons from $\nu A$ interactions? These neutrons
could be captured by the LSD metal structures and
produce gamma-ray photons. Given that the range of
a gamma-ray photon with a typical energy of 8~MeV in
iron is $\sim 4$~cm, such gamma-ray photons can freely
escape from the detector structures and produce a signal in the scintillation counters. This possibility needs
to be carefully checked.

A significant uncertainty in the parameters of the
neutrinos (especially their energy) emitted during the
core collapse of a rapidly rotating SN in the absence of
convincing numerical simulations makes an attempt to
solve the inverse problem meaningful. That is we can
attempt to answer the following question: if all five
LSD pulses are attributable to the interaction of neutrino
radiation from SN1987A, then what characteristics
should it have possessed? This question can be
answered only through a direct simulation of the neutrino
interaction with the LSD material and surrounding
soil.

The goal of our study is to simulate the LSD
response to neutrino radiation from SN 1987A within
two possible scenarios:
\begin{itemize}
\item ``standard'' collapse with a mean neutrino energy
$\langle E\rangle_{\WI{\tilde{\nu}}{e}}\!\sim 15$~MeV;
\item the rotating collapsar model with two neutrino
signals. The first consists predominantly of electron
neutrinos with high energies $\langle E\rangle_{\WI{\nu}{e}}\!\sim 40$~MeV, while the
second is close in its parameters to the standard scenario
\cite{Imshennik1995}.
\end{itemize}

The neutrino energy characteristics in the rotational
mechanism were considered, for example, in
\cite{ImshMol2009, ImshMol2010}. It was shown that the typical energy could
indeed reach 40~MeV under certain conditions. To elucidate
the sensitivity of the results obtained to this specific
value, we additionally considered the case with an
intermediate energy of 30~MeV.

The paper is structured as follows. First we
describe LSD and then the neutrino–detector material
interaction channels. Next, we provide the characteristics
of the unique LSD event at 2:52~UT. Subsequently,
we describe the detector response simulation
method. The simulation results are presented in the
form of energy spectra for the main reactions and in a
summary table for the interactions of different types of
neutrinos with energies of 15, 30, and 40~MeV. In conclusion,
we discuss the results for both SN1987A
explosion scenarios under study.

Everywhere below we will call a single energy
release in the detector possessing all of the characteristic
signatures of a neutrino interaction a ``pulse'' and
a set of such pulses close in time an ``event''.

\section{LSD DESCRIPTION}
The liquid scintillation detector (LSD) that operated
since 1984 was designed to detect neutrinos from
stellar collapses \cite{Badino1984}. It was built in collaboration with
the Institute for Nuclear Research of the Academy of
Sciences of the USSR (presently the Institute for
Nuclear Research of the Russian Academy of Sciences)
and the Institute of Cosmogeophysics of the
Italian National Research Council in the tunnel under
Mont Blanc at a depth of 5200~m~w.e. Such a soil
thickness above the detector provides a reduction in
the cosmic-ray muon flux by six orders of magnitude.
The 12-km-long road tunnel linking Italy and France
passes from the southeast to the northwest.

LSD consisted of 72 scintillation counters $1.0 \times 1.5 \times 1.0$~m$^3$ each. The counter box was made of 0.4-cm thick
stainless steel. The counters formed three three-level
sections (columns) in the form of a parallelepiped
with an area of $6 \times 7$~m$^2$ and a height of 4.5~m (see
Figs. 4--6 below).

Iron containers placed on one another and holding
two counters each comprised the detector structure.
The walls of the containers had different thicknesses in
order that there be a 2-cm layer of iron between the
adjacent faces of the counters in the column. The side
walls facing the corridors and the bottoms of the containers
had a thickness of 2 cm. Thus, given the 0.4-cm
thickness of the counter walls, the scintillator of the
counters in the column was separated by 2.8-cm layers
of iron. The layer of iron was 2.8~cm between the scintillators
of the neighboring columns and 4.4~cm
between the rock soil and the scintillator. The detector
stood on a 10-cm-thick iron platform. To reduce the
influence of natural soil radioactivity, the chamber
walls were coated with steel plates with a total weight
of about 114~t.

There was a space 6 cm in height above and below
the detector where it was intended to place two layers
of resistive streamer tubes for experiments on muon
physics, but with the appearance of the NUSEX project
in 1982 \cite{NUSEX} capable of solving these problems, it
was decided to abandon the resistive streamer tubes.

The scintillator based on white spirit $\isn{C}{}_n\isn{H}{}_{2n}$, $\langle n\rangle \approx 9.6$,
contained PPO (1~g/L) and POPOP (0.03~g/L).
Each counter was viewed by three FEU-49B photomultiplier
tubes (PMTs) with a photocathode diameter
of 15~cm. The total signal of three PMTs produced
by approximately 15 photoelectrons corresponded to
an energy release of 1~MeV in the counter. The energy
releases in the counter were analyzed when the signals
from three PMTs coincided with a resolution time of
200~ns.

The energy threshold of the 16 internal counters
better shielded from the background was $\WI{E^{\mathrm{in}}}{HET} = 5$~MeV; the detection threshold for the external
56 counters was $\WI{E^{\mathrm{ex}}}{HET} = 7$~MeV. A pulse from the coincidence
circuit of any of the 72 counters was a trigger
for the entire detector. In this case, the pulse amplitude
(with an energy release above $\WI{E}{LET} = 0.8$~MeV)
and time in a time window of 500~$\mu s$ were recorded in
each of the 72 counters.

The energy calibration of the counters was performed
based on the peaks from the energy releases of
atmospheric muons (175~MeV for vertical muons) and
gamma-ray photons from the captures of neutrons
(spontaneous \isn{Cf}{252} fission products) by hydrogen
($E_\gamma = 2.2$~MeV) and nickel ($E_\gamma \approx 9$~MeV) nuclei.

The energy resolution of a scintillation counter is
described by the formula
\begin{equation}
\frac{\sigma}{E}=\frac{1}{2.35}\Big[0.26\pm\sqrt{\frac{0.31}{E(\mbox{MeV})}+0.055}\Big],
\end{equation}
It is based on the determination of the resolution
$\eta=\triangle E/E$, $\triangle E=2.35\sigma$, $\sigma=1/\sqrt{\WI{N}{pe}}=1/\sqrt{a E}$, where
$\triangle E$ is the full width at half maximum of the resolution
function (quasi-Gaussian), $\sigma$ is the confidence interval
(68\%
measurement error), $\WI{N}{pe}$ is the number of
photoelectrons on three PMTs, and $a$ is the number of
photoelectrons on three PMTs from an energy release
of 1~MeV. The first term in parentheses represents the
scatter of the light collection coefficient as a function
of the burst location in the counter; the first and second
terms of the sum under the radical are determined
by the fluctuations in the number of photoelectrons
$\WI{N}{pe}$ and the PMT gain unbalance, respectively.

\section{SELECTION OF CANDIDATES
FOR SIGNAL DETECTION}
The detection system of the LSD experiment
allowed one to detect electron antineutrino ($\WI{\tilde{\nu}}{e}$)–scintillator
interaction products: positrons from the reaction
$\WI{\tilde{\nu}}{e}+p \rightarrow n + e^+$ (the Reines–Cowan IBD reaction)
and gamma-ray photons from the capture of a
neutron by protons $n+p\rightarrow d+\gamma$ or iron nuclei $n + \isn{Fe}{} \rightarrow \isn{Fe}{} + \sum\gamma$. This sequence of reactions gives the
main antineutrino detection ``signature''.

If a positron is detected, then the amplitude is
equal to the sum of the positron kinetic energy and the
$e^+e^-$ annihilation gamma-ray photon energy
($\sim 1$~MeV).

The particle detection system is determined by the
upper, $\WI{E}{HET}$, and lower, $\WI{E}{LET}$, threshold functions and
the edge effect \cite{Amanda}.

The mean counting rate of detector trigger pulses is
about 45 per hour. The mean number of neutron-like
pulses in a time window of 500~$\mu s$  is 0.1 and 0.8 for the
internal and external counters, respectively.

A neutrino burst is identified by the appearance of
a series of pulses with an amplitude greater than $\WI{E}{HET}$
in a time $t < 20$~s. The detector background can imitate
a true event. The background imitation frequency was
estimated in an online detector data analysis,
\begin{equation}
\WI{F}{im}=m\frac{(mt)^{k{-}1}}{(k{-}1)!}\exp(-mt),
\end{equation}
where $m$ is the frequency of background pulses, $t$ is the
pulse packet duration, and $k$ is the number of pulses.
The information about a series of pulses with a low
imitation frequency (less than $1/(5 \div 50)$ years) was
sent to printout from the detector computer.

The procedure for the selection of pulse candidates
for neutrino signals from gravitational collapses is
described in detail in \cite{Dadykin1987, Aglietta1989}.

\subsection{Neutrino–LSD Material Interaction Reactions}
Apart from the IBD reaction, LSD was able to
detect particles from the interaction of neutrinos with
carbon and iron nuclei in the detector scintillator and
structure. The reactions and detection thresholds are
given in Tables 1 and 2. The detector can record the
delayed pulses (gamma-ray photons and neutrons)
from the deexcitation of \isn{Mn}{}, \isn{Fe}{} and \isn{Co}{} nuclei owing
to the low threshold $\WI{E}{LET}$, which ``opens'' all counters
for reading after the arrival of a trigger, a pulse higher
than $\WI{E}{HET}$. The pulses from these gamma-ray photons
and the gamma-ray photons from the captures of neutrons
by the detector iron in the reactions of $\WI{\tilde{\nu}}{e} (\WI{\nu}{e})$
with the detector material and surrounding soil can
have energies up to 15~MeV and can also be recorded
by the detector as single triggers.
\begin{figure*}[htb]
\tabl[1]{
Charged-current (CC) neutrino interaction reactions
}
\begin{center}
\begin{tabular}{|c|c|c|}
\hline
Type of neutrinos & Nucleus: \isn{C}{} & Nucleus: \isn{Fe}{}\\
\hline
& $\WI{\tilde{\nu}}{e}+\isn{C}{12}\rightarrow\isn{B}{12}+e^+$ $(\WI{E}{th}=14.4~\mbox{MeV})$ & $\WI{\tilde{\nu}}{e}+\isn{Fe}{56}\rightarrow\isn{Mn}{56}^*+e^+$ $(\WI{E}{th}=12.5~\mbox{MeV})$\\
$\WI{\tilde{\nu}}{e}$& $\isn{B}{12}\rightarrow\isn{C}{12}+e^-+\WI{\tilde{\nu}}{e}$& $\isn{Mn}{56}^*\rightarrow\isn{Mn}{56}+\gamma$\\
& & $\isn{Mn}{56}^*\rightarrow\isn{Mn}{55}+n$\\
& & $\isn{Mn}{56}^*\rightarrow\isn{Cr}{55}+p$\\
\hline
& $\WI{\nu}{e}+\isn{C}{12}\rightarrow\isn{N}{12}+e^-$ $(\WI{E}{th}=17.3~\mbox{MeV})$ & $\WI{\nu}{e}+\isn{Fe}{56}\rightarrow\isn{Co}{56}^*+e^-$ $(\WI{E}{th}=10~\mbox{MeV})$\\
$\WI{\nu}{e}$& $\isn{N}{12}\rightarrow\isn{C}{12}+e^++\WI{\nu}{e}$& $\isn{Co}{56}^*\rightarrow\isn{Co}{56}+\gamma$\\
& & $\isn{Co}{56}^*\rightarrow\isn{Co}{55}+n$\\
& & $\isn{Co}{56}^*\rightarrow\isn{Fe}{55}+p$\\
\hline
\end{tabular}
\end{center}
\end{figure*}
\begin{figure*}[htb]
\tabl[2]{
Neutral-current (NC) neutrino interaction and electron scattering (ES) reactions
}
\begin{center}
\begin{tabular}{|c|c|c|c|}
\hline
Type of neutrinos & Nucleus:: \isn{C}{} & Nucleus:: \isn{Fe}{} & $e^-$\\
\hline
$\WI{\tilde{\nu}}{e}$ & $\WI{\nu}{i}+\isn{C}{12}\rightarrow\WI{\nu}{i}+\isn{C}{12}^*$  & $\WI{\nu}{i}+\isn{Fe}{56}\rightarrow\WI{\nu}{i}+\isn{Fe}{56}^*$  &\\
$\WI{\nu}{e}$ & $(\WI{E}{th}=15.1~\mbox{MeV})$ & $(\WI{E}{th}=15~\mbox{MeV})$ & $\WI{\nu}{i}+e^-\rightarrow\WI{\nu}{i}+e^-$\\
$\WI{\tilde{\nu}}{\mu,\tau}$& $\isn{C}{12}^*\rightarrow\isn{C}{12}+\gamma$ & $\isn{Fe}{56}^*\rightarrow\isn{Fe}{56}+\gamma$ & \\
$\WI{\nu}{\mu,\tau}$& $\isn{C}{12}^*\rightarrow\isn{C}{11}+n$ & $\isn{Fe}{56}^*\rightarrow\isn{Fe}{55}+n$ & \\
& $\isn{C}{12}^*\rightarrow\isn{B}{11}+p$ & $\isn{Fe}{56}^*\rightarrow\isn{Mn}{55}+p$ & \\
\hline
\end{tabular}
\end{center}
\end{figure*}

\section{THE EVENT RECORDED BY LSD DURING
THE SN1987A EXPLOSION}
A series of five pulses was recorded in the LSD
experiment in 7~s on February 23, 1987, at 2~h 52~min
Universal Time (UT). Information with signal characteristics
was sent to the International Astronomical
Union (IAU) on February 28 \cite{IAUC4323, Aglietta1987}. Below, the pulses
near 2:52~UT will be called the first signal.

The Kamiokande II collaboration (Japan, USA)
reported the detection of neutrinos from SN 1987A in
the IAU on March 6 and prepared a publication \cite{Hirata}.
The team of the IMB experiment (USA) sent the same
report to the IAU on March 11 and prepared a publication
on March 12 \cite{Bionta}. The group of the Baksan
Underground Scintillation Telescope (BUST) \cite{Alekseev}
reported the extraction of a neutrino signal from their
data on April 6. With the corrections for the clock
accuracy, all three signals occurred simultaneously at
7:35~UT. LSD recorded two pulses at 7:35~UT \cite{Aglietta1987}. The
pulses near 7:35~UT will be called the second signal.

The technique of the experiments of both KII and
IMB Cherenkov detectors and BUST scintillation
telescope did not allowed one to identify precisely the
$\WI{\tilde{\nu}}{e}p$ interactions: the detectors could not detect an np
capture and, thus, in contrast to LSD, had no signature
of this reaction.

The KII signal containing 11 pulses (initially 12)
was interpreted by the authors as the one produced by
$\WI{\tilde{\nu}}{e}p$ interactions based on approximate isotropy of the
direction of particle tracks in the pulses. The authors
explained the fact that the particle tracks in 7 of the 11
pulses were directed to the forward hemisphere by
fluctuations. The situation with the IMB signal is similar:
7 of the 8 tracks were also directed to the front
hemisphere. The KII authors assumed that 2 of the 11
pulses in their signal could be produced by $\nu e$ scattering,
but such a number of $\nu e$ interactions led to a contradiction
with the theoretical predictions about the
total energy of the neutrino flux \cite{LoSecco1987,LoSecco1989,Malgin1998}.

Thus, the recorded LSD cluster contained 5 pulses
in the time interval of 7 s bounded by the first and last
pulses (Fig. 1).
\begin{figure}[htb]
\begin{center}
\includegraphics[width=0.6\textwidth]{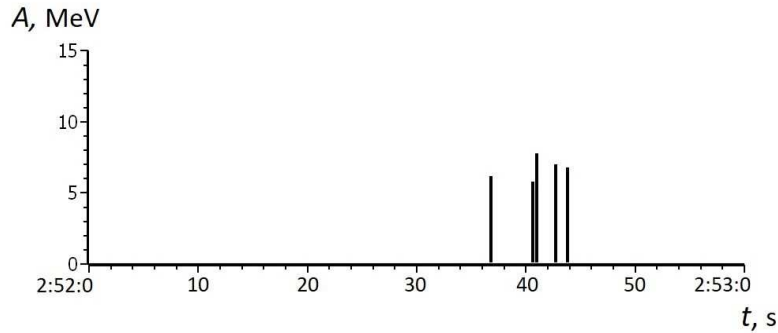}
\end{center}
\vspace{-0.4cm}
\caption{The series of pulses recorded in the LSD (2 h 52 min
UT)}
\end{figure}
The cluster candidates followed at LSD with a frequency
of about 0.5 per day, but this cluster had an
exceptionally low background imitation frequency $\WI{F}{im} \sim 1.8\times 10^{-3}$ per day and appeared for the first time
since the beginning of the detector operation in January
1985. A cluster close in imitation frequency ($\sim 4\times 10^{-3}$) was recorded only on April 27, 1986; it consisted
of 11 pulses in 67~s \cite{Agafonova2013}.

The occurrence times of the pulses in LSD and
their amplitudes (refined after additional calibrations
\cite{Aglietta1989}]) are given in Table 3. Only one of the five measured
pulses was accompanied by a neutron-like pulse
offset from the trigger signal in the detector by 278~$\mu s$
with an energy of 1.4~MeV.
\begin{figure*}[htb]
\tabl[3]{
Characteristics of the recorded LSD cluster
}
\begin{center}
\begin{tabular}{|c|c|c|c|}
\hline
Counter number & Pulse time, & Amplitude, & Neutron-like \\
   & hour.min.sec  &  MeV  & pulse \\
\hline
31 external &	2.52.36.792	& 6.2	& ---\\
14 internal	& 2.52.40.649	& 5.8	& ---\\
25 external	& 2.52.41.007	& 7.8	& +1.0 after 278~$\mu s$\\
35 internal	& 2.52.42.696	& 7.0	& ---\\
33 internal &	2.52.43.800 &	6.8	& ---\\
\hline
\end{tabular}
\end{center}
\end{figure*}

For information, Table 4 presents the times and the
numbers of pulses in the IMB \cite{Hirata}, KII \cite{Bionta}, and BUST
\cite{Alekseev} detectors. Such a small number of pulses was
caused not only by the distance to the source ($\sim 50$~kpc
from the Earth), but also by the relatively small sizes of
the detectors.
\begin{figure}[htb]
\tabl[4]{
The signals recorded by the IMB, KII, and BUST
detectors
}
\begin{center}
\begin{tabular}{|c|c|c|}
\hline
 & First–second  & Number of \\
Detector & pulse times, & pulses\\
 & hour.min.sec &\\
\hline
IMB	&7.35.35 -- 7.35.47	& 12\\
\hline
KII &	7.35.41 -- 7.35.44 &	8\\
\hline
BUST & 7.36.06 -- 7.36.21 &	6\\
\hline
\end{tabular}
\end{center}
\end{figure}

\section{SIMULATION OF THE LSD RESPONSE}
The simulation was performed using the \textsc{Geant4}
software package of version 10.3 \cite{Geant4}. This software
package allows detailed simulations of the passage of
elementary particles through matter to be carried out
by the Monte Carlo method. \textsc{Geant4} also has the necessary
tools for specifying objects of complex geometry
and includes a wide set of theoretical models describing
the interaction of elementary particles with matter.
A detailed description of the set of physical models
applied in our simulations can be found in \cite{Manukovskiy2016}. It
should be noted that these physical settings were specially
selected and optimized to simulate the experimental setups in underground low-background laboratories.
The testing and debugging were performed on
a series of experimental data, from relatively simple
experiments on the interaction of protons and $\pi$-mesons of fixed energy with iron and lead targets to
more complex experimental setups with liquid-scintillator-based detectors at the Artemovsk Scientific Station
of the Institute for Nuclear Research in plaster (25~m~w.e.) and salt (316 and 570~m~w.e.) mines. The
results obtained in our numerical simulations agreed
well with the results of experiments \cite{Lom2015}.

There are no models to describe the neutrino–
matter interaction in the \textsc{Geant4} library package.
Therefore, information about the processes involving
neutrinos should be added to the simulation from outside.
We used standard formulas (see, e.g., \cite{BurrThom2004}) to
describe the following processes: (1) neutrino capture
by a nucleus, (2) inelastic neutrino scattering by a
nucleus, and (3) electron scattering. The main problem
here is to calculate the amplitude and distribution
of Gamow–Teller and Fermi resonances in the daughter
nucleus, which dominate in these processes on
nuclei at the typical neutrino energies under consideration.
For a full calculation we need information on
the neutrino interaction both in the detector scintillator
and in the LSD protective metal structures and the
soil surrounding the detector. Below we present a
compilation of the bibliographic data on the elements
(with isotopes) involved: \isn{H}{1} \cite{Vogel1984, BurrThom2004, StrumVis2003}, \isn{C}{12,13} \cite{Yoshida2008, DapoPaar2012, Kolbe1999, Suzuki2011, Fukugita1988}, \isn{O}{16,18} \cite{Suzuki2011, Kolbe2002, Kuram1990, Haxton1987, Lazauskas2007, Anderson1983}, \isn{Al}{27} \cite{Stetcu2004, Fujita1999}, \isn{Si}{28} \cite{Luttge1996, Anderson1991}, \isn{Cr}{50,52{-}54} \cite{Petermann2007, Langanke2004, Muto1985, Nabi2016}, \isn{Fe}{54,56,57} \cite{DapoPaar2012, Paar2008, Lazauskas2007, Toivanen2001, Bandy2017, Caurier1999, Kolbe1999, Joud2005}, \isn{Ni}{58,60} \cite{Joud2005, Suzuki2011, Caurier1999}.

\begin{figure}[htb]
\begin{center}
\includegraphics[width=0.6\textwidth]{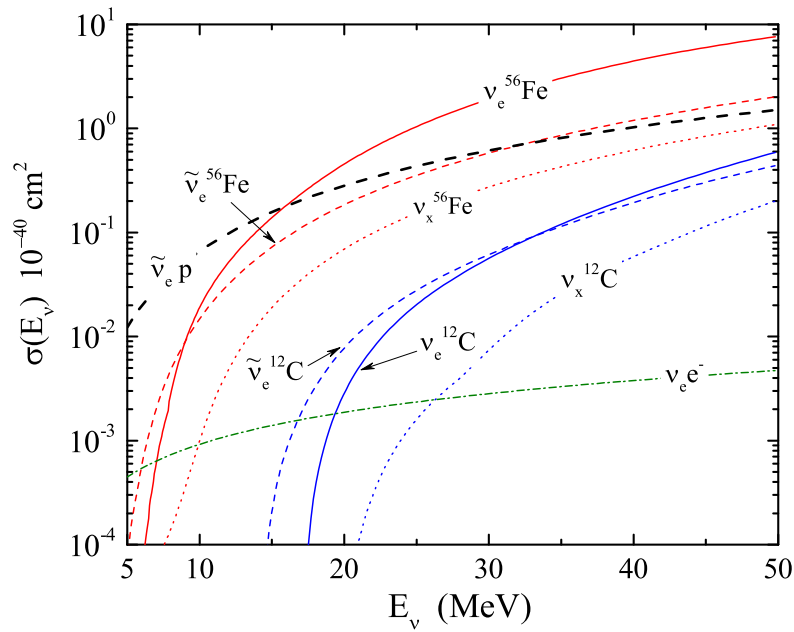}
\end{center}
\vspace{-0.4cm}
\caption{Cross sections for the neutrino interaction with several
selected targets as a function of energy}
\end{figure}
Figure 2 presents the calculated cross sections for
the interaction of neutrinos and antineutrinos with
several most important targets: hydrogen and carbon
nuclei (the scintillator composition is $\isn{C}{}_n \isn{H}{}_{2n}$), iron
(the protective and bearing structures of the detector),
and electrons. The solid and dashed lines indicate the
results for the interaction of electron neutrinos and
antineutrinos via the charged channel, respectively.
The dotted and dash–dotted lines indicate inelastic
scattering and electron scattering, respectively. As can
be seen, the cross section for the reaction $\WI{\nu}{e} + \isn{Fe}{56}$
exceeds the cross section for the main IBD reaction
already at about 15~MeV. This is particularly important
for the rotational SN explosion mechanism under
consideration \cite{ImshRya2004}, which can produce high-energy
neutrinos.

To demonstrate the uncertainty with regard to the
calculated cross sections available in the literature,
Table 5 gives the cross sections for the neutrino interaction
with an iron nucleus $\WI{\nu}{e} + \isn{Fe}{56} \rightarrow \isn{Co}{56} + e^-$ at
neutrino energy $E_\nu = 40$~MeV, according to some
works. As can be seen, the available data allow us to
find the cross sections to within the accuracy of factor 2. This
should be taken into account in the subsequent analysis
of our results. Note, however, that the main detection
channel, the electron antineutrino–hydrogen
interaction reaction, does not contain such uncertainties.
\begin{figure}[htb]
\tabl[5]{
Cross section for the neutrino interaction with an
iron nucleus at $E_\nu = 40$~MeV
}
\begin{center}
\begin{tabular}{|c|c|}
\hline
Reference & $\sigma_\nu, 10^{-40}$ cm$^2$ \\
\hline
\cite{Kolbe2001} & 3.72\\
\hline
\cite{Lazauskas2007} & 5.41\\
\hline
\cite{Paar2008} & 2.1\\
\hline
\cite{Bandy2017} & 3.04\\
\hline
\cite{Ryazh2018} & 4.2\\
\hline
This paper & 4.3\\
\hline
\end{tabular}
\end{center}
\end{figure}

\subsection{LSD Design and the Surrounding Soil}
Particular attention and much effort were given to
the creation of the LSD geometry. Where possible, it
was necessary to include all of the structural elements
that could affect the numerical simulation results in
the model. The simulated geometry also took into
account the granite surrounding the detector. The
granite of the Mont Blanc range is composed mainly
of silicon dioxide $\isn{Si}{} \isn{O}{}_2$. The chemical composition
of the granite used in our computations and the mass
fractions of the main chemical elements are shown in
Tables 6 and 7.
\begin{figure}[htb]
\begin{minipage}{0.49\textwidth}
\tabl[6]{
Chemical composition of granite
}
\begin{tabular}{|c|c|}
\hline
Compound &	Mass fraction, \% \\
\hline
$\isn{Si}{}\isn{O}{}_2$	&72.76 \\
\hline
$\isn{Al}{}_2\isn{O}{}_3$	&13.96 \\
\hline
$\isn{K}{}_2\isn{O}{}$ 	&4.35\\
\hline
$\isn{Na}{}_2\isn{O}{}$	&3.76\\
\hline
$\isn{Fe}{}_2\isn{O}{}_3$	&2.18\\
\hline
$\isn{Ca}{}\isn{O}{}$	&1.09\\
\hline
$\isn{H}{}_2\isn{O}{}$	&0.99\\
\hline
$\isn{Mg}{}\isn{O}{}$	&0.65\\
\hline
$\isn{Ti}{}\isn{O}{}_2$	&0.26\\
\hline
\end{tabular}
\end{minipage}
\begin{minipage}{0.49\textwidth}
\tabl[7]{
Mass fractions of chemical elements
}
\begin{tabular}{|c|c|}
\hline
Element	& Mass fraction, \% \\
\hline
\isn{O}{}	&49.23\\
\hline
\isn{Si}{}	&34.01\\
\hline
\isn{Al}{}	&7.39\\
\hline
\isn{K}{} 	&3.61\\
\hline
\isn{Na}{}	&2.79\\
\hline
\isn{Fe}{}	&1.53\\
\hline
\isn{Ca}{}	&0.78\\
\hline
\isn{Mg}{}	&0.39\\
\hline
\isn{Ti}{}	&0.16\\
\hline
\isn{H}{}	&0.11\\
\hline
\end{tabular}
\end{minipage}
\end{figure}

Since LSD is well shielded from various
background sources, the only neutrino–granite
nuclei interaction products that could reach the inner
detector volume and lead to signal detection by its
counters are neutrons. Therefore, in our preliminary
computations we studied the penetrability of neutrons
in the granite. As our simulation for various initial
neutrino energies (from 1 to 20~MeV) showed (see
Fig. 3), the maximum penetration depth does not
exceed 1.5~m, and only about 5\%
of the neutrons traverse
a distance more than 100~cm. Therefore, in our
computations the granite layer sensitive to the interaction
with neutrinos had a thickness of 1~m. In contrast,
the total thickness of the granite surrounding the room
with LSD and the adjacent segment of the road tunnel
in the numerical geometry was 3~m.
\begin{figure}[htb]
\begin{center}
\includegraphics[width=0.4\textwidth]{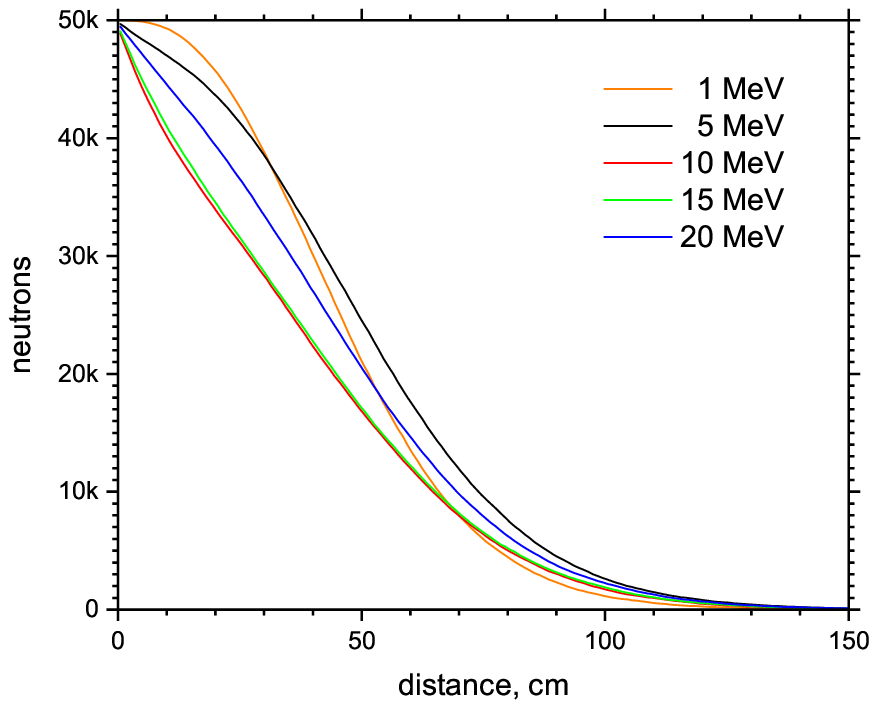}
\includegraphics[width=0.4\textwidth]{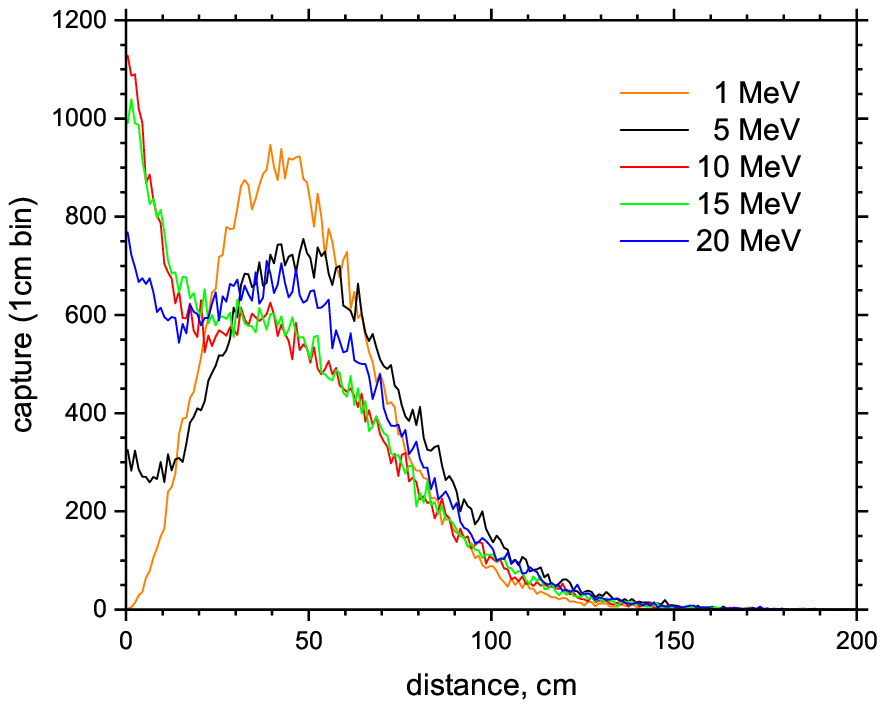}
\end{center}
\vspace{-0.4cm}
\caption{Left: change in the number of neutrons as one
recedes from the source point in the granite. Right: the distribution
of neutron capture points in the granite as a function
of distance.}
\end{figure}

A general view of the geometry for LSD with an
inner detecting volume (counters in iron containers),
outer protective steel plates, and a granite layer around
the experimental room, and the tunnel fragment is
presented in Fig.~4. The internal structure of the
detector surrounded by a 1-m granite layer is shown in
more detail in Figs. 5 and 6.
\begin{figure*}[htb]
\begin{center}
\includegraphics[width=.6\textwidth]{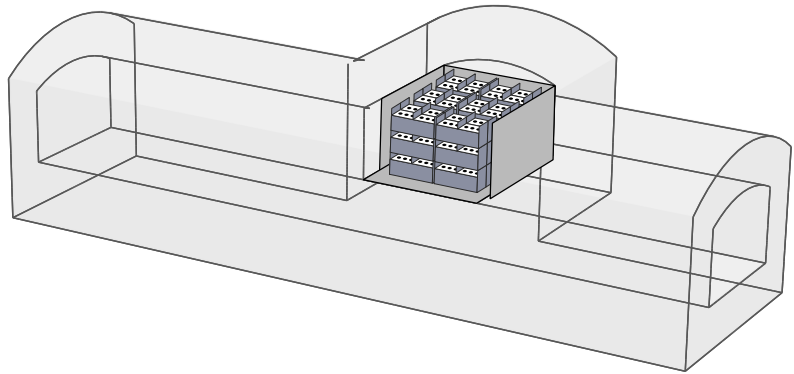}
\caption{General view of the numerical LSD geometry used
in our simulations}
\end{center}
\end{figure*}
\begin{figure}[htb]
\begin{minipage}{0.4\linewidth}
\includegraphics[width=.99\linewidth]{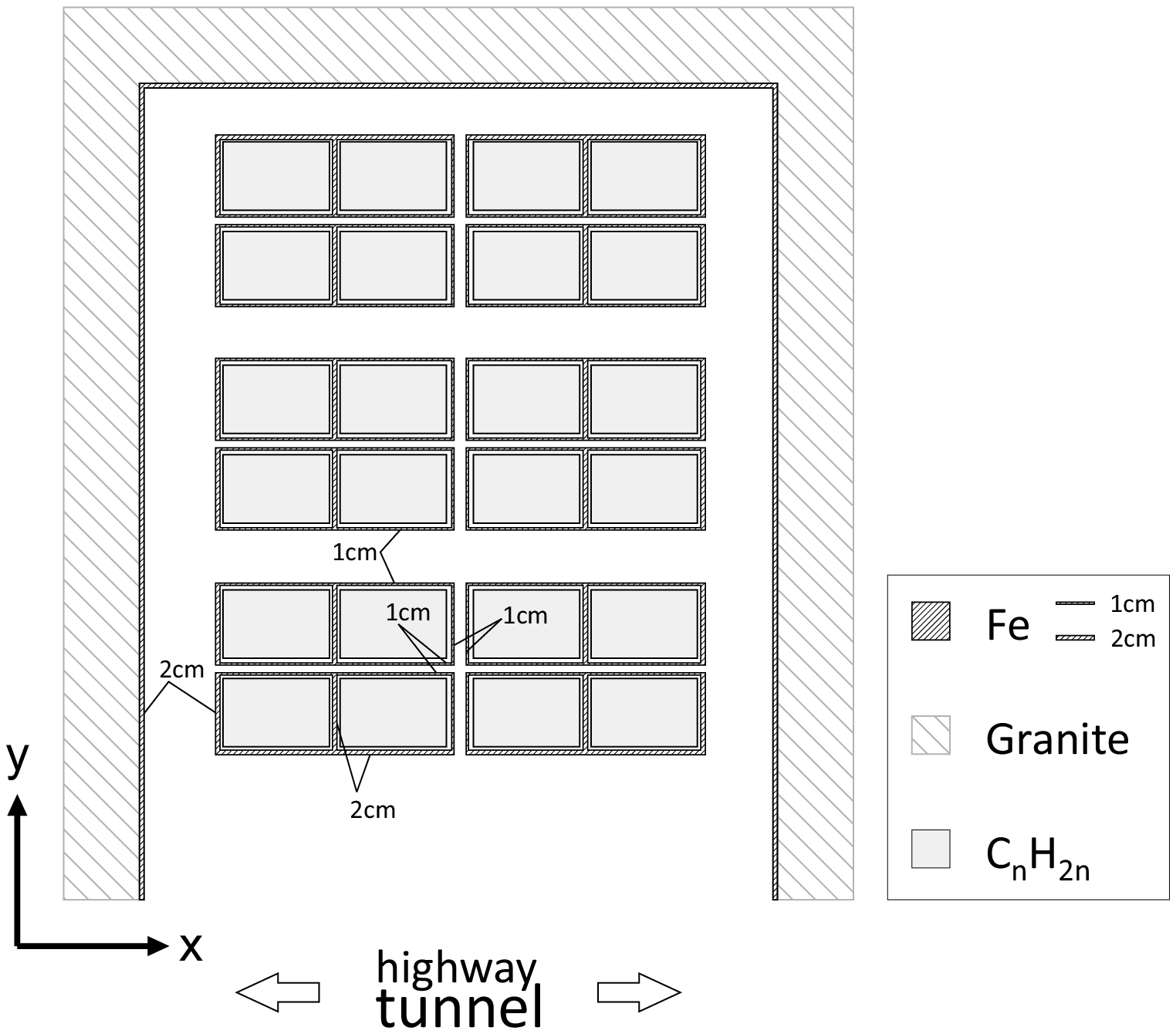}
\caption{LSD geometry (top view)}
\end{minipage}
\begin{minipage}{0.59\linewidth}
\includegraphics[width=.99\linewidth]{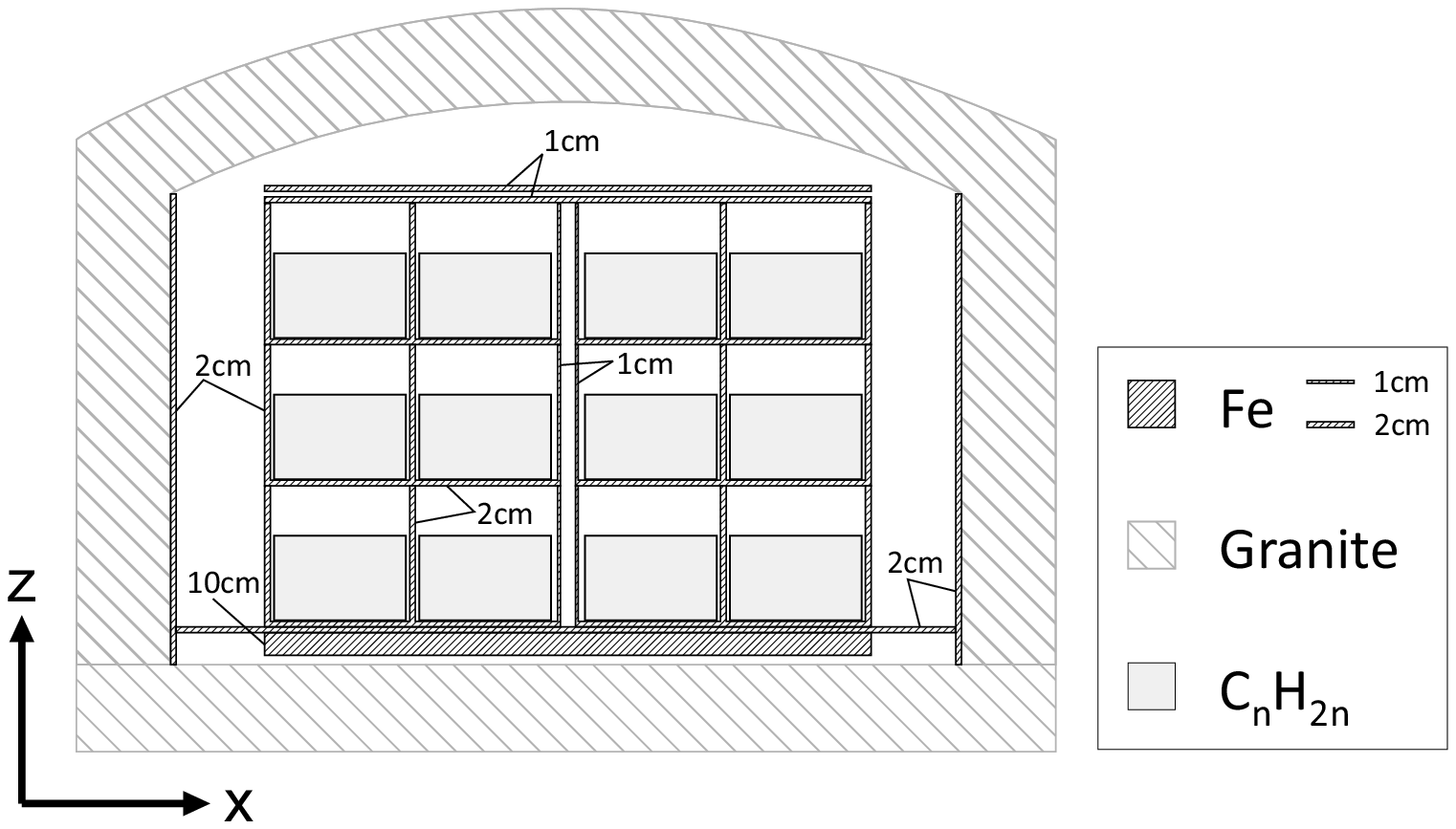}
\caption{LSD geometry (a view from the road tunnel)}
\end{minipage}
\end{figure}
In our computations the detector geometry is a set
of separate volumes with a complex shape that are created
through logical operations from a set of elementary
geometric figures (cylinders, cones, parallelepipeds,
etc.). Homogeneous chemical composition and
density are specified within a single volume. The precomputed
cross sections for the reactions of neutrinos
with various nuclei (see above) were used to simulate
the interaction of neutrino radiation with the LSD
structural elements.

\subsection{Calculation of the Neutrino Interaction
with the Detector Elements}
Consider some volume of the experimental setup
geometry with index $i$, total mass $M^i$, and chemical
composition specified by the mass fractions $X^i_{A,Z}$ of the
individual nuclei in its chemical composition.
Here, $Z$ and $A$ are the nuclear charge and mass number,
respectively. The number of neutrino interactions
via a given channel of reactions $q$ with a nucleus $(A,Z)$
in the volume $i$ under consideration per unit flux of
neutrinos of a given type is then
\begin{equation}
N^i_{A,Z}=\frac{M^i}{\WI{m}{u}}\frac{X^i_{A,Z}}{A}\sigma^q_{A,Z}(E_\nu),\label{NiAZ}
\end{equation}
where $\sigma^q_{A,Z}$ is the cross section for the corresponding
reaction, and $\WI{m}{u}$ is the atomic mass unit. As is easy to
understand, the total number of reactions (again per
unit flux) of neutrinos with energy $E_\nu$ via channel $p$
with the matter of a given volume will be
\begin{equation}
N^i=\sum\limits_{A,Z}N^i_{A,Z}=\frac{M^i}{\WI{m}{u}}\sum\limits_{A,Z}\frac{X^i_{A,Z}}{A}\sigma^q_{A,Z}(E_\nu),\label{Ni}
\end{equation}
while the total number of interactions with the detector
as a whole is defined by the obvious expression
\begin{equation}
\WI{N}{tot}=\sum\limits_{i}N^i.\label{Ntot}
\end{equation}
The probability that the reaction will occur in volume
$i$ is given by the ratio of Eqs. (\ref{Ni}) and (\ref{Ntot}). A similar
simulation of the nucleus with which the neutrino
interacts follows next in our computation. The probability
of the reaction with a nucleus $(A,Z)$ is equal to
the ratio of (\ref{NiAZ}) to (\ref{Ni}). The reaction products (daughter
nuclei, possibly, in an excited state, electrons/positrons,
etc.) are used as primary particles in our computation
using \textsc{Geant4}. The results obtained are normalized
to the total number of neutrino reactions with the
simulated detector geometry that is calculated based
on a specified neutrino flux. For example, for monochromatic
neutrinos with energy $E_\nu$ from SN1987A,
provided that the total energy released isotropically in
the form of such neutrinos was $\WI{E}{tot}$, the number of
interactions can be estimated from the formula
\begin{equation}
\WI{N}{int}\approx 1.2\frac{\WI{E}{tot}}{10^{53}~\mbox{erg}}
\frac{10~\mbox{MeV}}{E_\nu}
\frac{\WI{M}{tot}}{100~\mbox{t}}
\left(\!\frac{51.4~\mbox{kpc}}{\WI{R}{SN}}\!\right)^2\!\!
\sum\limits_{A,Z,q}\!\frac{X_{A,Z}}{A}\frac{\sigma^{\mathrm q}_{A,Z}(E_\nu)}{10^{-42}\mbox{cm}^2},
\end{equation}
where $\WI{R}{SN}$ [kpc] is the distance to the SN. Generally,
however, it is necessary to integrate over the spectrum
of incoming neutrinos.

The data acquisition and signal detection in the
counters during our numerical simulation were performed
by a technique similar to the operation of the
recording equipment in an experiment.

\section{RESULTS}
\subsection{Simulation of the LSD Response
to Neutrino Fluxes from the Soil}
At the first stage of our simulation we investigated
the assumption about the connection of the signal
recorded with LSD at 2:52~UT with neutrons that
could be produced in the reactions of the interaction
of neutrinos with the granite that surrounded the
detector.

First, we studied the penetrability of neutrons in
the soil (see Fig.~3 above) to estimate the effective volume
(and mass) of the granite that could affect the
detector. The penetration depth was $\sim100$~cm for the
entire range of possible neutron energies. It should be
noted that the neutrons still need to traverse more than
4 cm of the iron shield after their escape from the granite.
Nevertheless, the granite volume in which the
neutrons that are able to escape into the room with the
experimental setup can be produced is quite impressive.
The granite mass in this volume is more than
3000~t, which exceeds the mass of all iron and steel
structures by almost an order of magnitude.

Second, we simulated the detector response to
neutrons coming from the granite. These simulations
also showed a weak dependence of the results on the
initial neutron energy. As an example, we provide the
energy distribution of the signals recorded by the LSD
counters for two initial energies, 1 and 8~MeV (see
Fig.~7). As can be seen, the results agree satisfactorily
in energy with the experimental data, which makes the
assumption \cite{Yen} about the possible role of the granite
in explaining the recorded signal very plausible. However,
despite the large mass of the granite that surrounded
the detector, our simulations of the interaction
with neutrino radiation showed that neutrons are
a very rare product in all of the possible reactions of
neutrinos with nuclei in the granite, which does not
allow the number of recorded pulses to be explained at
all (see Table~8 below).

\begin{figure}[htb]
\includegraphics[width=.49\textwidth]{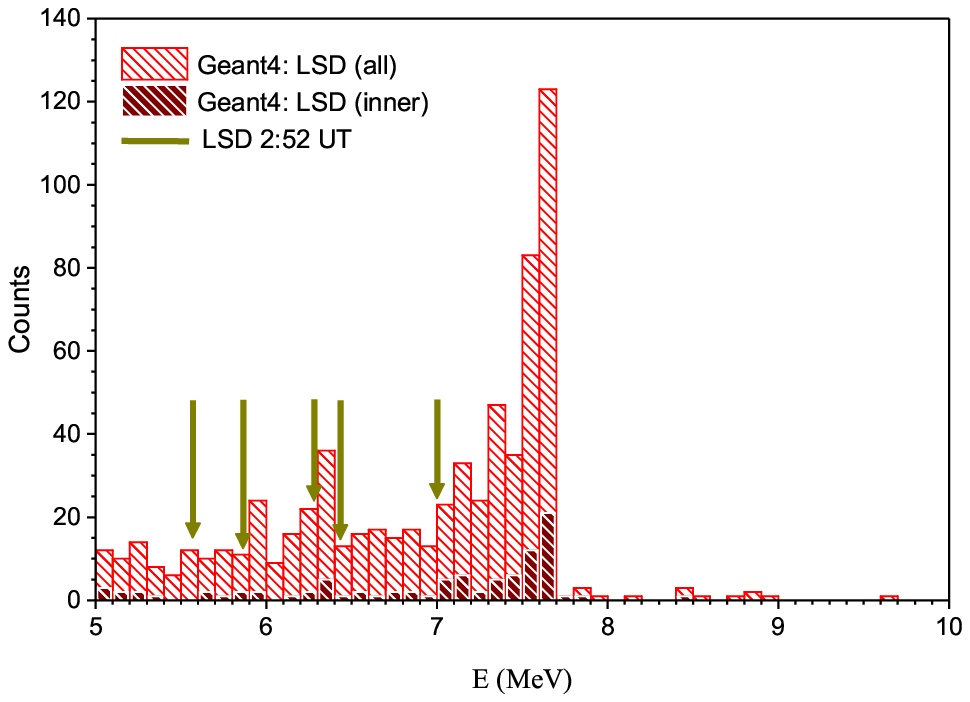}
\includegraphics[width=.49\textwidth]{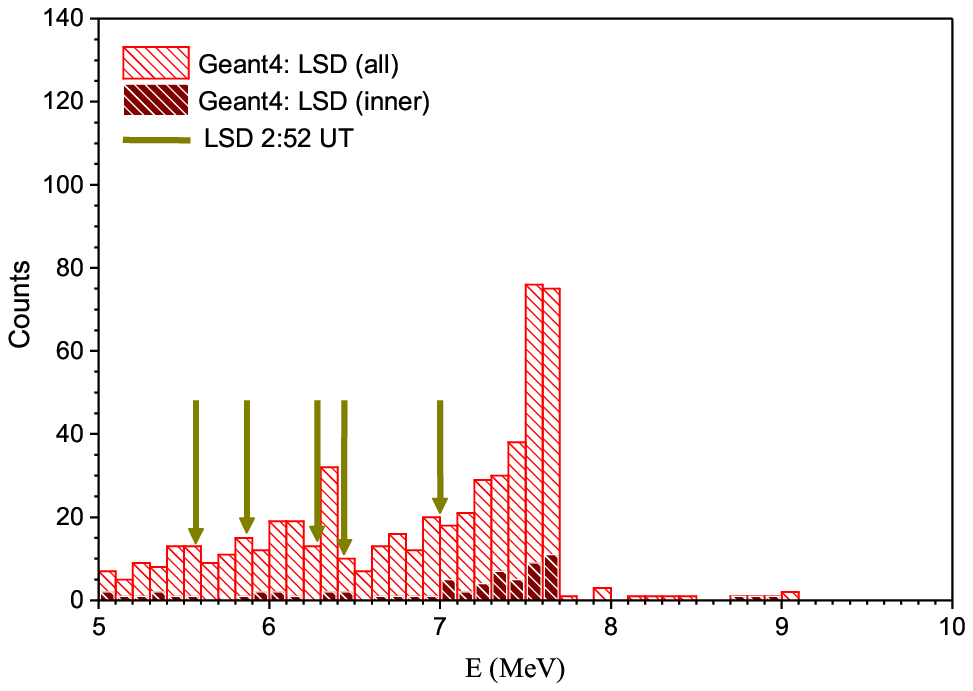}
\vspace{-0.6cm}
\caption{Spectrum of energy releases (with an energy above
5 MeV) in the LSD counters for the neutrons with an initial
energy of 1 (left) and 8 (right) MeV produced in the granite. The
energy distribution for the internal and all detector counters
is indicated by the dark and light colors, respectively.
The arrows indicate the energies of the signals recorded in
LSD at 2:52~UT}
\end{figure}

\subsection{Simulation of the LSD Response
to the Neutrino Burst}
We performed a series of simulations using the
technique described above for various neutrino energies,
types, and reaction channels. During the simulation
the detector and its surrounding soil were irradiated
by a neutrino flux of specific composition (with
an energy that was fixed or distributed over the equilibrium
spectrum). The total energy radiated by the
SN in the form of neutrinos in all our simulations was
$10^{53}$~erg. Figure~8 presents the simulation results for a
flux of electron antineutrinos with an energy distributed
over the equilibrium spectrum,
\begin{equation}
F_\nu(E_\nu)\propto E_\nu^3\exp\left(-\frac{3E_\nu}{\langle E_\nu\rangle}\right),
\end{equation}
with a mean value of $\langle E_{\nu}\rangle = 15$~MeV for the charged current
reaction channel. The spectrum of energy
releases in the counters has a maximum at energies
near 7–8~MeV and a slowly decaying ``tail''. We simulated
50000 reactions of neutrinos with specified properties.
As expected, most of the reactions that led to
the detection of signals with an energy above 5~MeV in
the scintillation counters occurred in the scintillator
itself, while the main nucleus that entered into the
reactions with neutrinos was hydrogen (note that the
scale in upper panels of Fig.~8 is logarithmic).
\begin{figure}[htb]
\begin{center}
\includegraphics[width=0.7\linewidth]{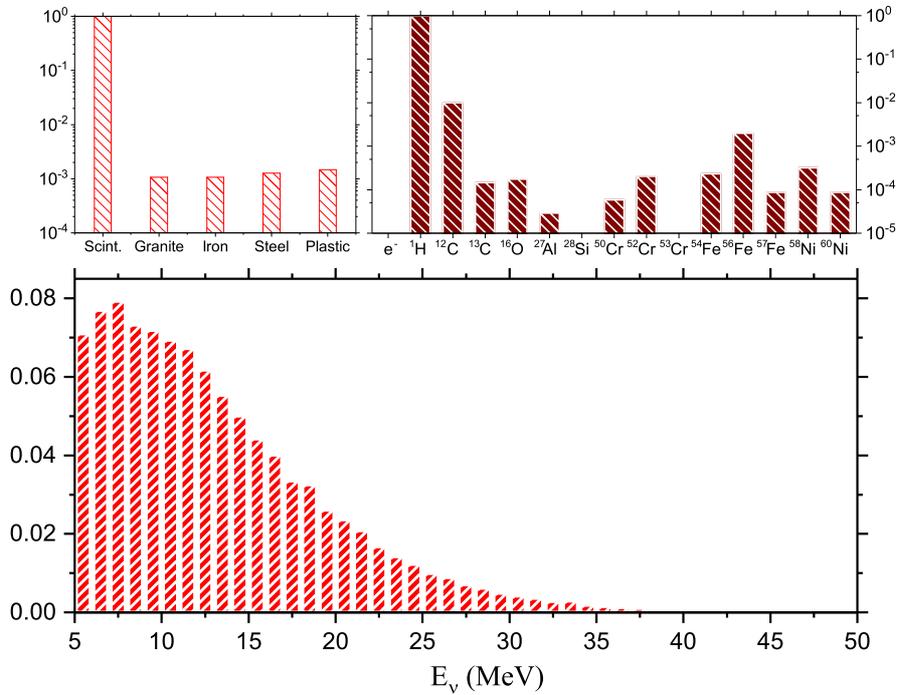}
\end{center}
\vspace{-0.6cm}
\caption{The relative number of reactions in various materials
(upper left) and on various nuclei (upper right) accompanied by the detection
of a signal with an energy above 5 MeV. Bottom panel: the normalized
spectrum of energy releases in the counters for
electron antineutrinos with an energy distributed over the
equilibrium spectrum with a mean value of 15~MeV in the
charged-current reactions}
\end{figure}

Figure 9 presents the same data, but for a flux of
electron neutrinos with an energy distributed over the
equilibrium spectrum with a higher mean energy of 40~MeV (see \cite{ImshRya2004}) in the charged-current reactions. The
spectrum of energy releases in this case turns out to be
fairly extended (approximately to 50~MeV) and monotonically
decaying. Neutrinos mostly react with the
material of the iron and steel detector structures, but a
comparable number of reactions occur in the scintillator
as well. Iron and carbon isotopes are the nuclei that
enter into the reactions with neutrinos most often.
\begin{figure}[htb]
\begin{center}
\includegraphics[width=0.7\linewidth]{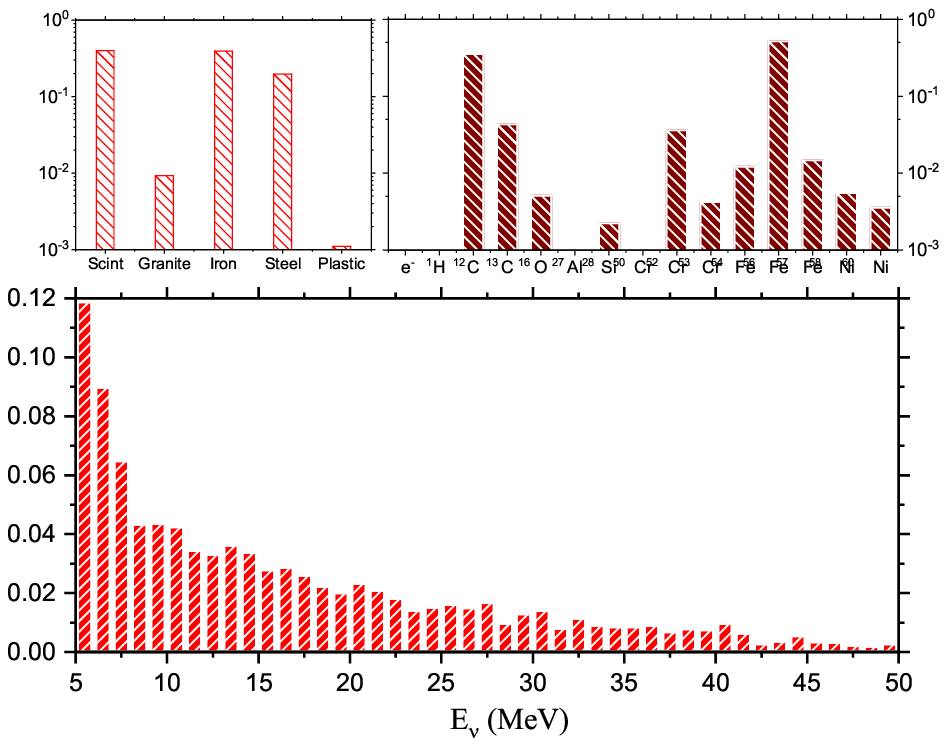}
\end{center}
\vspace{-0.6cm}
\caption{Same as Fig.~8, but for electron neutrinos with an
energy distributed over the equilibrium spectrum with a
mean value of 40~MeV in the charged-current reactions.}
\end{figure}

Finally, Fig. 10 shows the simulation results for the
channel of neutral-current and electron scattering
reactions. The neutrino spectrum is an equilibrium
one with a mean energy of 40~MeV. The spectrum of
energy releases has a pronounced maximum in the
energy range 5–10~MeV, closely corresponding to the
energies recorded in the experiment. Nevertheless, it is
still worth noting that there are signals with high energies
(up to $\sim 30$~MeV) in the spectrum as well. As in the
previous case, the iron, steel, and scintillator play a
major role. However, apart from the iron and carbon
isotopes, the reactions of neutrino scattering by electrons
contribute noticeably to the number of recorded
pulses.
\begin{figure}[htb]
\begin{center}
\includegraphics[width=0.7\linewidth]{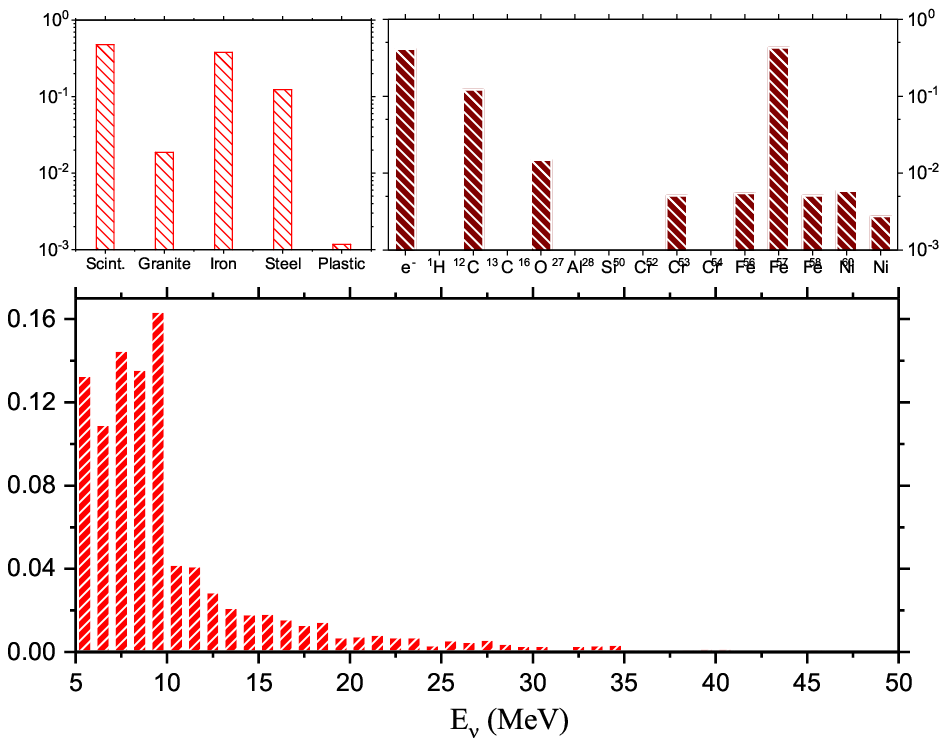}
\end{center}
\vspace{-0.6cm}
\caption{Same as Fig.~8, but for electron neutrinos with an
energy distributed over the equilibrium spectrum with a
mean value of 40~MeV in the charged-current reactions}
\end{figure}

Table 8 summarizes the most important results of
our numerical simulations for a wide set of neutrino
energies and types of reactions. The distance to the
neutrino radiation source was set equal to the distance
to SN1987A. At the same time, the neutrino flux was
assumed to be isotropic, while its total energy in each
simulation was $10^{53}$~erg. Table~8 gives the expected
(from the simulation results) number of pulses in
LSD, the probability that a given energy release fell
into the range 5–10~MeV, the fraction of pulses from
the reactions in the granite, and the fraction of pulses
from the reactions among the products of which there
were neutrons.
\begin{figure*}[htb]
\tabl[8]{
Summary table of the simulation results for various types of reactions and energies of neutrinos from SN 1987A
}
\begin{center}
\begin{tabular}{|c|c|c|c|c|c|}
\hline
Type & Mean &	Total number &	Probability that &	Fraction of  &	Fraction of pulses\\
 of	& neutrino  & of expected   & one pulse fell into &	reactions in &	with neutron \\
reaction & energy, MeV  & pulses & range 5–10 MeV &	soil, $\times10^{-2}$ & production, $\times10^{-2}$
\\
\hline
\hline
 & 15 &	1.92 &	0.379	& 0.11 &	0.06\\
\cline{2-6}
$\WI{\tilde{\nu}}{e}$ &30 &	5.6 &	0.175 &	0.29 &	0.92\\
\cline{2-6}
 & 40	& 8.46	& 0.135 &	0.54 &	1.96\\
 \hline
 \hline
 & 15 &	0.08	& 0.635	& 0.17 &	1.10\\
 \cline{2-6}
$\WI{\nu}{e}$ & 30	& 1.1	& 0.449 &	0.45 &	2.43\\
 \cline{2-6}
& 40 &	2.6 &	0.371 &	0.93 &	2.86\\
\hline
\hline
 &15 &	0.046 &	0.69 &	0.06	& 0.68\\
 \cline{2-6}
 $NC{+}ES$ & 30	& 0.22 &	0.69 &	0.92 &	1.64\\
 \cline{2-6}
 & 40 &	0.41 &	0.69	& 1.93 &	3.52\\
 \hline
\end{tabular}
\end{center}
\end{figure*}

As can be seen from the last two columns
of the table, the contribution of the granite and
neutrons to the number of recorded pulses from the
neutrino radiation is negligible for all of the possible
energies and reaction channels. This is apparently
because the granite surrounding LSD is composed
mostly of $\alpha$-particle nuclei (\isn{O}{16}, \isn{Si}{28}, see Table~7) with
high binding energies and neutron separation energies.
Out of the still produced neutrons, only a small fraction
will be able to escape from the soil (see Fig.~3) and
an even smaller fraction will be able to traverse the
LSD protective structures to give a signal in the counters.
Thus, unfortunately, the elegant idea \cite{Yen} about
the influence of the surrounding soil on the signal
from the SN detected in LSD does not work.

As can be seen from Table 8, the main factor that
increases the total number of recorded events is a high
energy. The neutrino scattering reactions contribute
insignificantly to the total number of events at all of
the possible energies. On the other hand, the spectrum
of energy releases observed in the experiment is most
closely reproduced precisely in the scattering reactions.

\subsection{Errors in the Results Obtained}
The quantities in Table~8 are affected by the errors
in determining some characteristics of the detector
and the neutrino flux. The scatter of parameters of different
\textsc{Geant4} versions also contributes to the systematic
error of our simulations.

To reduce the systematic error of our simulations,
we performed a sufficiently large number of simulations.
For each set of parameters the number of reactions
with neutrinos was $5\times 10^4$. This made it possible
to bring the statistical error for the main quantities
obtained in our simulations (the total number of
expected events and the probability of falling into the
energy range 5–10~MeV) to fractions of percent. The
values of the quantities in the last two columns of
Table~8 (the fractions of reactions in the soil and of
events with neutron production) were obtained with a
lower accuracy, because these events are fairly rare.
However, the statistical error did not exceed 20\%
even for them.

The example with \isn{Fe}{56} (Table~5) gives an idea of the
systematic errors when including a set of neutrino
interaction cross sections in the simulations. As can be
seen, the difference in cross sections is of the order of
a factor of 2 and this quantity depends strongly on the
energy and type of nucleus. However, the cross section
for the main $\WI{\tilde{\nu}}{e}p$ reaction is well known.

The error associated with the masses of the materials
used in our simulations and the geometry of the
detector and the experimental room contribute no
more than 12\%
to the systematic error of the results
obtained.

As experience shows, the result also depends to
some extent on the \textsc{Geant4} version used \cite{Lom2015}. This is
related to the details of the physics and the parameters
of the processes included by the developers in the software
package version. In our simulations we used the
\textsc{Geant4} version 10.3.0.

The main uncertainty is associated with the neutrino
burst characteristics used: the total burst energy
($\sim 10^{53}$~erg), the assumption that the neutrino radiation
is spherically symmetric, the energy characteristics of
the neutrino fluxes, etc. In the real case of SN1987A,
they can differ from those used in the simulations.
However, this uncertainty can be partly corrected,
depending on the implied SN model: for example,
using a different total energy leads simply to a renormalization
of the derived numbers of pulses in the
detector etc.

\section{DISCUSSION}
Let us discuss the results obtained and compare
them with the experimental data. For these purposes,
we will use Fig.~11.
\begin{figure}[htb]
\begin{center}
\includegraphics[width=0.7\linewidth]{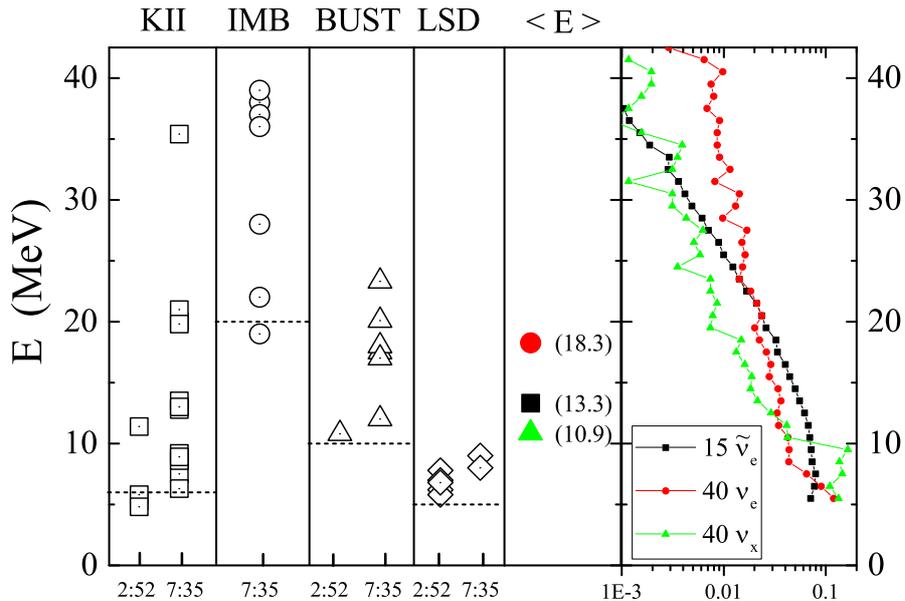}
\end{center}
\vspace{-0.6cm}
\caption{Compilation of experimental data from four detectors:
KII (squares), IMB (circles), BUST (triangles), and
LSD (diamonds). Only the recorded energies are shown.
For all of the detectors, except IMB, the first and second
data sets correspond to 2:52 and 7:35~UT, respectively.}
\end{figure}
It presents a compilation of experimental
data from four detectors \cite{Dadykin1989}: KII (squares), IMB (circles), BUST (triangles), and LSD (diamonds). Only
the recorded energies are shown. For all detectors,
except IMB, the first and second data sets correspond
to (approximately) 2:52 and 7:35~UT, respectively.
IMB did not ``see'' the first signal, in contrast to the
remaining detectors (for a discussion of the situation
with the non-recognition of the reality of the first signal,
see \cite{Rujua1987}). The spectra of energy releases in LSD
according to our simulations (Figs.~8–10) and the corresponding
mean energies $\langle E\rangle$ (the numerical value in
MeV is given in parentheses) are also shown. The horizontal
dashed lines indicate the threshold level for
each detector. Here, it is pertinent to recall that the
notion of a detection threshold is conventional.

During the second signal at 7:35~UT the number of
events in the LSD was 2, in excellent agreement with
the value calculated by us for standard collapse (15-MeV antineutrinos) --- 1.92 (Table~8).

During the first signal no detector, except LSD,
was able to record the signal from electron neutrinos
\cite{ImshRya2004}. The fact that KII and BUST still ``saw'' something
(Fig.~11) suggests that there was a slight admixture
of low-energy electron antineutrinos in the first
signal (there was no pulse at IMB due to a high, $\sim 20$~MeV, threshold). This means that we can also attribute
one of the five pulses in LSD to the detection of the
IBD reaction, the one that (alone) was accompanied
by a characteristic signature of this event, see Table~3.
We assign an energy of 15~MeV to the neutrino that
generated this pulse based on the spectrum of energy
releases in KII and BUST and the absence of a signal
in IMB (Fig.~11).

Let us now consider the main ingredient of the
rotational model—high-energy (40~MeV) electron
neutrinos. We see that the combination of the numbers
of charged-current (2.6) and neutral-current
(0.41) pulses gives 3 events, which, given all of the
available uncertainties, coincides remarkably with $5 -
1 = 4$ pulses from electron neutrinos. The 40-MeV
muon and tau neutrinos at the time of the first signal
in the rotating collapsar model can originate from
electron neutrinos due to oscillations.

Thus, the combination of events at the time of the
first signal in LSD can appear as:
\begin{equation}
5 = 1(15~\mbox{MeV}~\WI{\tilde{\nu}}{e}) + 1(40~\mbox{MeV}~\WI{\nu}{x}) + 3(40~\mbox{MeV}~\WI{\nu}{e}).\label{5equal}
\end{equation}
Naturally, this is only one of the possible variants,
through apparently the most probable one: we cannot
increase the fraction and energy in $\WI{\tilde{\nu}}{e}$, otherwise other
detectors should have recorded much more pulses. In
contrast, the neutral-current reactions give a too small
number of events due to their small cross section.

However, we see a major problem in the measured
energies of the events: all of the energy releases in LSD
(in both first and second signals) lie in a narrow range,
from 5~MeV (threshold) to approximately 9~MeV. Let
us calculate the probability that all five pulses of the
combination (\ref{5equal}) considered above fell into this range
using Table~8:
\begin{equation}
P(5) = 0.379\times 0.69\times(0.371)^3 \approx 0.013.
\end{equation}

Turning to Fig.~11, we see that the signal in LSD
indeed breaks out of the general series. In all of the
remaining detectors the pulse energy distribution is
wide, as it must be (see the mean expected pulse energies
in LSD in the same figure). We see that the signal
in LSD agrees neither with the computed spectra of
energy releases (Figs.~8–10) nor with the pulse energies
in other detectors. All five pulses of the first signal
in LSD have low energies (see also Table 3), but high energy
neutrino ($\WI{\nu}{e}$ and/or $\WI{\nu}{\mu,\tau}$) fluxes are required to
explain their number (recall that the neutrino–matter
interaction cross section increases with energy
approximately as $\sigma_\nu\sim E_\nu^2$).

We can admit an error in the energy calibration of
the LSD counters, as a result of which the pulse amplitude
will be at least halved. This could explain the
energy spectrum of the pulses in LSD, which clearly
breaks out of both the spectra of other detectors and
the predictions of our simulations. However, this
explanation comes into conflict with the agreement
between the experimental and computed spectra of
muon energy losses (from 40 to 400~MeV) in the
counter and with the correspondence of the calibration
based on the 2.2-MeV ($\gamma$-ray photons from
$np$-captures) and 9-MeV ($\gamma$-ray photons from $n\isn{Ni}{}$ captures)
peaks. Unfortunately, the LSD calibration can
no longer be checked in the range 10–40~MeV of interest
to us (for example, this could have been done with
the LINAC calibration system used in Super-Kamiokande
\cite{Migenda2020}), because LSD was dismantled in 1998.

The closeness of the LSD pulse amplitudes to the
detection threshold $\sim 7$~MeV may suggest a ``background'' origin of the signal. The connection of the
change in the background of the experiment (the
number of radionuclide decays in the soil surrounding
the detector and its materials) with the SN 1987A
explosion is discussed in \cite{Malgin_GW2020}.

Nevertheless, the facts that confirm a high significance
of the LSD signal remain:
\begin{itemize}
\item among the LSD signal candidates for the detection
of a $\nu$--burst selected over 14 years of operation,
this signal has the lowest background imitation probability
 \cite{Ryazhskaya2006};
\item the probability of a chance coincidence of the
LSD signal with the SN1987A explosion is extremely
low, less than $<1.4\times 10^{-6}$ \cite{Galeotti2016,Aglietta1991};
\item the LSD signal enters into the central part of the
unique and unexplainable 6-hour event in the interval
approximately from 1 to 7 hours UT on February 23,
1987, formed by the set of experimental data from four
neutrino detectors and two gravitational wave detectors
in Rome and Maryland that operated during the
SN1987A explosion \cite{Malgin_GW2020,Galeotti2016}.
\end{itemize}

\section{CONCLUSIONS}
In several aspects the derived quantities are
approximate. Nevertheless, based on a detailed simulation
of the interactions of neutrinos from the gravitational
stellar core collapse with the LSD material
and surrounding soil, we can draw the following conclusions.

First, our simulation showed that the number of
recorded LSD pulses both at the time of the first burst
2:52~UT and at the time of the second one 7:35~UT
could indeed be obtained within the rotational SN
explosion mechanism. However, the energy characteristics
of both groups of pulses do not correspond to
the computed expected spectra of energy releases and
the data from other detectors.

Second, the hypothesis suggested in \cite{Yen} that the
signal in LSD at 2:52~UT could be generated by the
neutrinos produced by neutrino fluxes in the soil surrounding
the detector does not pass a quantitative
check. As our simulations showed, the spectra of such
pulses are indeed very similar to the LSD data. However,
their number is negligible for all of the neutrino
radiation parameters considered.

Above we presented several possible explanations
for this. At present, there is no explanation that would
satisfy all of the data, both theoretical and experimental
ones. Nevertheless, we hope that the results presented
here will serve as an important step in long-term
attempts to elucidate the nature of the unique signal
in LSD coincident in time with the SN1987A
explosion.

\section{ACKNOWLEDGMENTS}
This study was performed in 2019 and was essentially
finished in early 2020, before the death of our coauthors
A.S.~Malgin and O.G.~Ryazhskaya. We think their contribution
to this study to be decisive and devote the paper to
their memory with gratitude. We thank V.S.~Imshennik and
D.K.~Nadyozhin, who has also passed away, for the useful
discussions and sincere interest in this study. We are grateful
to the referee whose remarks contributed significantly to an
improvement of the text of our paper.

\section{FUNDING}
The work of A.V. Yudin was supported by the Russian
Science Foundation (project no. 21-12-00061).


\begin{references}
\bibitem{IAUC4316} IAUC 4316: 1987A, N. Cen. 1986. February 24, 1987
\bibitem{Aglietta1987} M. Aglietta et al., EuroPhys. Lett. 3 (1987) 1315

\bibitem{Alekseev} E.N. Alekseev et al., Sov. Phys. JETP Lett. 45 (1987) 461

\bibitem{Bionta} R.M. Bionta, et al., Phys. Rev. Lett. 58, 1494 (1987)

\bibitem{Hirata} K. Hirata, et al., Phys. Rev. Lett. 58, 1490 (1987)

\bibitem{MSK1} Ya. B. Zel’dovich and O. Kh. Guseinov, Sov. Phys.
Dokl. 10, 524 (1965)

\bibitem{MSK2} W.D. Arnett, Can. J. Phys. 44, 2553 (1966)

\bibitem{Rujua1987} A. De Rujula, Phys. Lett. B, Vol. 193, N 4, 514 (1987)

\bibitem{Berezinsky1988} V.S. Berezinsky, C. Castagnioli, V.I. Dokuchaev, P. Galeotti. Il Nuovo Cimento, 11C, N3, 287 (1988)

\bibitem{Imshennik1995} V.S. Imshennik, Space Sci. Rev. 74, 325(1995).

\bibitem{Drago2016} A. Drago, G. Pagliara, Europ. Phys. J. A, 52, 41, 15 pp (2016)

\bibitem{ImshRya2004} V. S. Imshennik and O. G. Ryazhskaya, Astron. Lett.
30, 14 (2004)

\bibitem{Yen} S. Yen, TRIUMF Vancouver, Canada (talk 18-Apr 2017)

\bibitem{ImshMol2009} V.S.~Imshennik V.O.~Molokanov, Astron. Lett., \textbf{35}, 12, 799–815 (2009)

\bibitem{ImshMol2010} V.S. Imshennik, V.O.~Molokanov, Astron. Lett., \textbf{36},  10,  721–737 (2010)

\bibitem{Badino1984} Badino G. et al. Nuovo Cimento (1984) 7C, 573

\bibitem{NUSEX} G. Battistoni et al., Phys. Lett. B, \textbf{133}, 6, p. 454-460

\bibitem{Amanda} A. Porta ``\textit{Energy measurement in LVD to reconstruct Supernova neutrino emission}''. PhD Thesis of Torino University (2005). P.159

\bibitem{Dadykin1987} V. L. Dadykin, G. T. Zatsepin, V. B. Korchagin, et al.,
JETP Lett. 45, 593 (1987).

\bibitem{Aglietta1989} Aglietta, M. et al., ``\textit{Search for neutrinos from collapsing stars at Mont Blanc}'' In ``Vulcano 1988, Proceedings, Frontier objects in astrophysics and particle physics''  (1989) Vol.19 pp.103-120

\bibitem{IAUC4323} IAUC 4323: 1987A February 24, 1987

\bibitem{LoSecco1987} LoSecco J.M., Proceedings of the Second International Symposium UP-87, Baksan, USSR 1987 (Nauka, Moskow) 1988, p. 100.

\bibitem{LoSecco1989} LoSeccco J.M., Phys. Rev. D, 39, 1013 (1989)

\bibitem{Malgin1998} A. Malgin, il Nuovo Cim. C, 21, 317-329 (1998)


\bibitem{Agafonova2013} N. Yu. Agafonova, A. S. Malgin, and V. Fulgione,
J. Exp. Theor. Phys. 117, 258 (2013)

\bibitem{Geant4} V.N. Ivanchenko (for Geant4 Collab.), Nucl. Instrum. Methods A 502, 666 (2003)

\bibitem{Manukovskiy2016} K. V. Manukovskii, O. G. Ryazhskaya, N. M. Sobolevsky,
and A. V. Yudin, Phys. At. Nucl. 79, 631 (2016).

\bibitem{Lom2015} K.V. Manukovskiy et al., Proceedings of the 16th Lomonosov Conference, p. 72, 2015

\bibitem{BurrThom2004} Burrows A., Thompson T.A. ``\textit{Neutrino-Matter Interaction Rates in Supernovae}''. In: Fryer C.L. (eds) Stellar Collapse. Astrophysics and Space Science Library, vol 302. Springer, Dordrecht(2004)

\bibitem{Vogel1984} P. Vogel, Phys. Rev. D 29, 9 (1984)

\bibitem{StrumVis2003} A. Strumia, F. Vissani, Phys. Lett. B, 564, 1-2, p. 42-54, (2003)

\bibitem{Yoshida2008} T. Yoshida, T. Suzuki, S. Chiba et. al., Astroph. Journal, 686, 448–466 (2008)

\bibitem{DapoPaar2012} H. Dapo, N. Paar, Phys. Rev. C 86, 035804 (2012)

\bibitem{Kolbe1999} E. Kolbe, K. Langanke, P. Vogel, Nucl. Phys. A 652 91-100 (1999)

\bibitem{Suzuki2011} T. Suzuki, Journal of Physics: Conference Series 321, 012041 (2011)

\bibitem{Fukugita1988} M. Fukugita, Y. Kohayama, K. Kubodera, Phys. Lett. B, 212, 2 (1988)

\bibitem{Kolbe2002} E. Kolbe, K. Langanke, P. Vogel, Phys. Rev. D 66, 013007 (2002)

\bibitem{Kuram1990} T. Kuramoto, M. Fukugita, Y. Kohyama and K. Kubodera, Nucl. Phys. A, 512, 711-736, (1990)

\bibitem{Haxton1987} W.C. Haxton, Phys. Rev. D, 36, 8, (1987)

\bibitem{Lazauskas2007} R. Lazauskas and C. Volpe, Nucl. Phys. A 792, 219 (2007)

\bibitem{Anderson1983} B.D. Anderson, A. Fazely, R.J. McCarthy et al., Phys. Rev. C 27, 4 (1983)

\bibitem{Stetcu2004} I. Stetcu, C.W. Johnson, Phys. Rev. C 69, 024311 (2004)

\bibitem{Fujita1999} Y. Fujita, H. Akimune, I. Daito et al., Phys. Rev. C 59, 1 (1999)

\bibitem{Luttge1996} C. Luttge, P. Neumann-Cosel et al., Phys. Rev. C, 53, 1, pp.127-130 (1996)

\bibitem{Anderson1991} B.D. Anderson, N. Tamimi et al., Phys. Rev. C, 43, 1 (1991)

\bibitem{Petermann2007} I. Petermann, G. Martinez-Pinedo, K. Langanke, E. Caurier, Eur. Phys. J. A 34, 319 (2007)

\bibitem{Langanke2004} K. Langanke, G. Martinez-Pinedo, P. von Neumann-Cosel, and A. Richter, Phys. Rev. Lett. 93, 202501 (2004)

\bibitem{Muto1985} K. Muto, H. Horie, Nucl. Phys. A, 440, 2, p. 254-273 (1985)

\bibitem{Nabi2016} J. Nabi, R. Shehzadi, M. Fayaz, Astroph. and Space Science, 361, 95, 17 (2016)

\bibitem{Paar2008} N. Paar, D. Vretenar and P. Ring, J. Phys. G: Nucl. Part. Phys. 35, 014058 (2008)

\bibitem{Toivanen2001} J. Toivanen, E. Kolbe, K. Langanke, G. Martinez-Pinedo, P. Vogel, Nucl. Phys. A 694, 395–408 (2001)

\bibitem{Bandy2017} A. Bandyopadhyay, P. Bhattacharjee, S. Chakraborty, K. Kar, S. Saha, Phys. Rev. D 95, 065022 (2017)

\bibitem{Caurier1999} E. Caurier, K. Langanke, G. Martinez-Pinedo, F. Nowacki, Nucl. Phys. A 653 439-452 (1999)

\bibitem{Joud2005} A. Juodagalvis, K. Langanke, G. Martinez-Pinedo et al., Nucl. Phys. A, 747, 87-108 (2005)

\bibitem{Kolbe2001} E. Kolbe and K. Langanke, Phys. Rev. C, 63, 025802 (2001)

\bibitem{Ryazh2018} O.G. Ryazhskaya and S.V. Semenov, Phys. Atom. Nucl., 81, 2, 262–265 (2018)

\bibitem{Dadykin1989} V. L. Dadykin, G. T. Zatsepin, and O. G. Ryazhskaya,
Sov. Phys. Usp. 32, 385 (1989)


\bibitem{Migenda2020} J. Migenda ``Supernova Model Discrimination with Hyper-Kamiokande'', PhD thesis, University of Sheffield, arXiv:2002.01649

\bibitem{Malgin_GW2020} N. Agafonova, A. Malgin, E. Fischbach, arXiv:2107.00265 [nucl-ex]


\bibitem{Ryazhskaya2006} O. G. Ryazhskaya, Phys. Usp. 49, 1017 (2006)

\bibitem{Galeotti2016} P. Galeotti, G. Pizzella, Eur. Phys. J. C 76, 426 (2016)

 \bibitem{Aglietta1991} M. Aglietta et al., il Nuovo Cim., C 14, 171-193 (1991)


\end{references}
\end{document}